\documentstyle[preprint,prl,aps,epsfig,amssymb]{revtex}
\begin{document}
\draft
\tighten

\title{Depletion forces near curved surfaces}

\author{R. Roth, B. G\"otzelmann, and S. Dietrich}
\address{Fachbereich Physik, Bergische Universit\"at Wuppertal, D-42097
  Wuppertal, Germany}

\maketitle
\begin{abstract}

Based on density functional theory the influence of curvature on the depletion
potential of a single big hard sphere immersed in a fluid of small hard
spheres with packing fraction $\eta_{s}$ either inside or outside of a hard
spherical cavity of radius $R_{c}$ is calculated. The relevant features of
this potential are analyzed as function of $\eta_{s}$ and $R_{c}$. There is a
very slow convergence towards the flat wall limit
$R_{c}\to\infty$. Our results allow us to discuss the strength of depletion
forces acting near membranes both in normal and lateral directions and to make
contact with recent experimental results.
\end{abstract}


\narrowtext

Many biological processes are controlled by the interactions of macromolecules
with cell membranes. Besides highly specific interactions of steric and
chemical nature there are also entropic force fields which are omnipresent but
whose actions depend only on gross geometrical features. These so-called
depletion forces arise because both the membranes and the macromolecules
generate excluded volumes for the small particles forming the
solvent. Although these forces have been discussed for biological systems for
many years \cite{Zimmerman97}, the simultaneous presence of many other forces
severely impedes the precise analysis of depletion forces in such systems. Therefore 
it  is highly welcome that dissolved colloidal particles can be tailored such that
they resemble closely the effective model of monodispersed hard spheres
confined by hard walls \cite{Crocker94}. This allows one to study the depletion 
forces {\em exclusively} and to compare them {\em quantitatively} with theoretical 
results. Once
a satisfactory level of understanding has been reached for these model systems
one can apply with confidence this knowledge to the much more complex
biological systems. Moreover, in colloidal suspensions themselves depletion
forces can be exploited to organize self-assembled structures \cite{Dinsmore96}. 
This dedicated use requires a detailed knowledge of the mechanisms of depletion
forces, too.

Along these lines in recent years there has been significant progress in
understanding the depletion forces between two big spheres and between a
single big sphere and a flat wall based on experiments
\cite{Kaplan94a,Kaplan94b,Sober95,Milling95,Ohshima97,Rudhardt98}, analytic results 
\cite{Mao95,Biben96,Goetz98a,Goetz98b}, and simulations \cite{Dickman97}. A generic
feature of membranes is, however, that they are not flat. The variation of the local
curvature leads to a new quality of the depletion forces in that they are no
longer directed only normal to the surface, as for a flat wall or a wall with
constant curvature, but that there is also a lateral component which promotes
transport along the membrane. So far there are no systematic theoretical
studies available which accurately predict this important curvature dependence
of the depletion forces. Based on the quantitatively reliable density
functional theory developed by Rosenfeld \cite{RF89} we compute the depletion
potential for a big sphere of radius $R_{b}$ either inside or outside of a
spherical cavity with radius $R_{c}$, filled with a solvent of hard spheres of
radius $R_{s}$, as function of $R_{c}$ and of the packing fraction of the small
spheres $\eta_{s}$ defined as the fraction of the total volume occupied by the 
small spheres\cite{comment}. This enables 
us to make contact with a recent experiment in which the curvature dependence of
depletion forces has been investigated by monitoring colloidal particles
enclosed in vesicles \cite{Dinsmore98}.

\begin{figure} 
\centering\epsfig{file=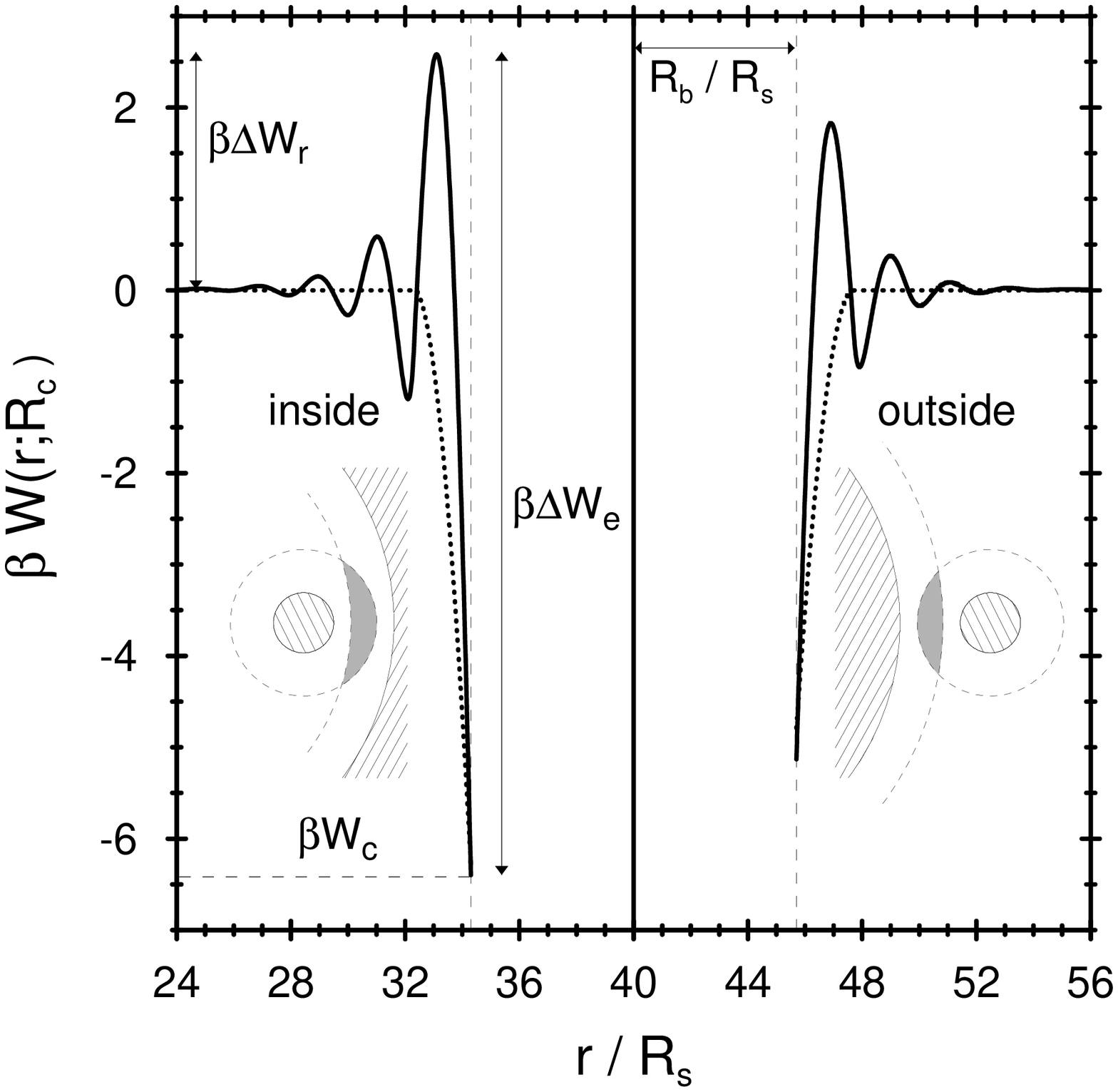,bbllx=10,bblly=60,bburx=550,bbury=580,width=13.5cm}

\vspace{0.5cm}

\caption{\label{fig:pot} Depletion potential $W(r;R_c)$ in units of
$k_{B}T=\beta^{-1}$ inside and outside of a spherical cavity centered at $r=0$
with radius $R_{c}=40 R_{s}$ (full vertical line). The radius of the big
sphere is $R_{b}=5.7 R_{s}$ and the packing fraction of the hard sphere
solvent (radius $R_{s}$) is $\eta_{s}=0.3$. Due to the hard core interactions
the center of the big sphere cannot enter the space indicated by the dashed
lines. The solid line is obtained from density functional theory whereas the
dotted line corresponds to the Asakura-Oosawa approximation which is valid for
small values of $\eta_s$ and exact in
the limit $\eta_{s}\to 0$. $W_{c}<0$, $\Delta W_{r}>0$, and 
$\Delta W_{e}=-W_{c}+\Delta W_{r}>0$ denote the depletion potential at contact with 
the wall and the potential barriers for reaching the wall and escaping from it,
respectively. The inset shows the different geometry of the (shaded) overlap of 
excluded volumes of a big sphere (hatched) inside and outside of a spherical 
cavity (hatched). (For reasons of clarity the schematic drawings correspond to
$R_b=0.7 R_s$ and $R_c=8.6 R_s$.)}
\end{figure}

Figure~\ref{fig:pot} shows our results for the depletion potential $W(r;R_c)$ in
units of $k_{B}T=\beta^{-1}$ inside and outside of a hard spherical cavity
centered at $r=0$. Here $W(r;R_c)$ is the difference of the grand canonical free 
energies of the hard sphere solvent in the presence and absence, respectively, 
of a big hard sphere whose center is fixed at a distance $r$ from the center of 
the cavity. The choice of the parameters corresponds to those of the experiment 
in Ref.~\cite{Dinsmore98}. The solid line denotes the actual depletion potential 
compared with its so-called Asakura-Oosawa approximation (AOa) \cite{Asakura54} 
(dotted line) which is valid only
for such small values of $\eta_{s}$ that the solvent can be treated as an
ideal gas and which is determined by the overlap of the excluded volumes
around the big sphere and the hard wall. The AOa, which has been used to
interpret the experimental data in Ref.~\cite{Dinsmore98}, satisfactorily predicts
the value $W_{c}$ of the depletion potential at contact and its derivation with
respect to $R_c$, but otherwise it fails
considerably. Whereas $W_{AOa}(r;R_c)$ is purely attractive and vanishes for
$r \lessgtr R_{c}\mp R_{b}\mp 2 R_{s}$, the actual potential is both attractive and repulsive
with the wavelength of the oscillations approximately given by $2 R_{s}$. The
correlation effects of the solvent generate the potential barrier $\Delta W_{r}$ 
(see Fig.~\ref{fig:pot}) which is completely missing within the AOa, and they
increase the range of the potential significantly beyond the AOa range of $2 R_s$.
The presence
of this potential barrier has very pronounced repercussions for the diffusion
dynamics of the big particle. The time $\tau$ required to overcome a potential
barrier $\Delta E$ is proportional to $\exp(\beta \Delta E)$. Therefore, with
$\Delta E=\Delta W_{r}$, for the big sphere it takes about $e^{2.5}\approx 12$
times longer to reach the wall from the center of the cavity or to escape from
the wall as compared with the time estimated from the AOa (see
Fig.~\ref{fig:pot}). This difference still awaits experimental confirmation.
For larger packing fractions $\eta_{s}$ this difference
becomes even larger. Thus the value $W_{c}=\Delta W_{r}-\Delta W_{e}$ of the
depletion potential at contact can be obtained from the ratio of rates of
escaping from the wall and reaching it, respectively. Taking into account only
the former one, as suggested by the AOa, would be misleading. The overall
structure of the depletion potential outside of the cavity is similar to the
one inside, but its amplitude is reduced (Fig.~\ref{fig:pot}). This is in
line with the trend predicted by the AOa which expresses the fact that the
overlap between the excluded volumes around a big sphere and the wall is
larger inside the spherical cavity than outside. This geometrical difference
between the excluded overlap volume inside and outside of the cavity is
schematically drawn as inset in Fig.~\ref{fig:pot}. In the inset the boundaries of
the excluded volumes around the big sphere and the cavity wall are indicated by
dashed lines and the overlap of excluded volumes is shaded.

\begin{figure} 
\centering\epsfig{file=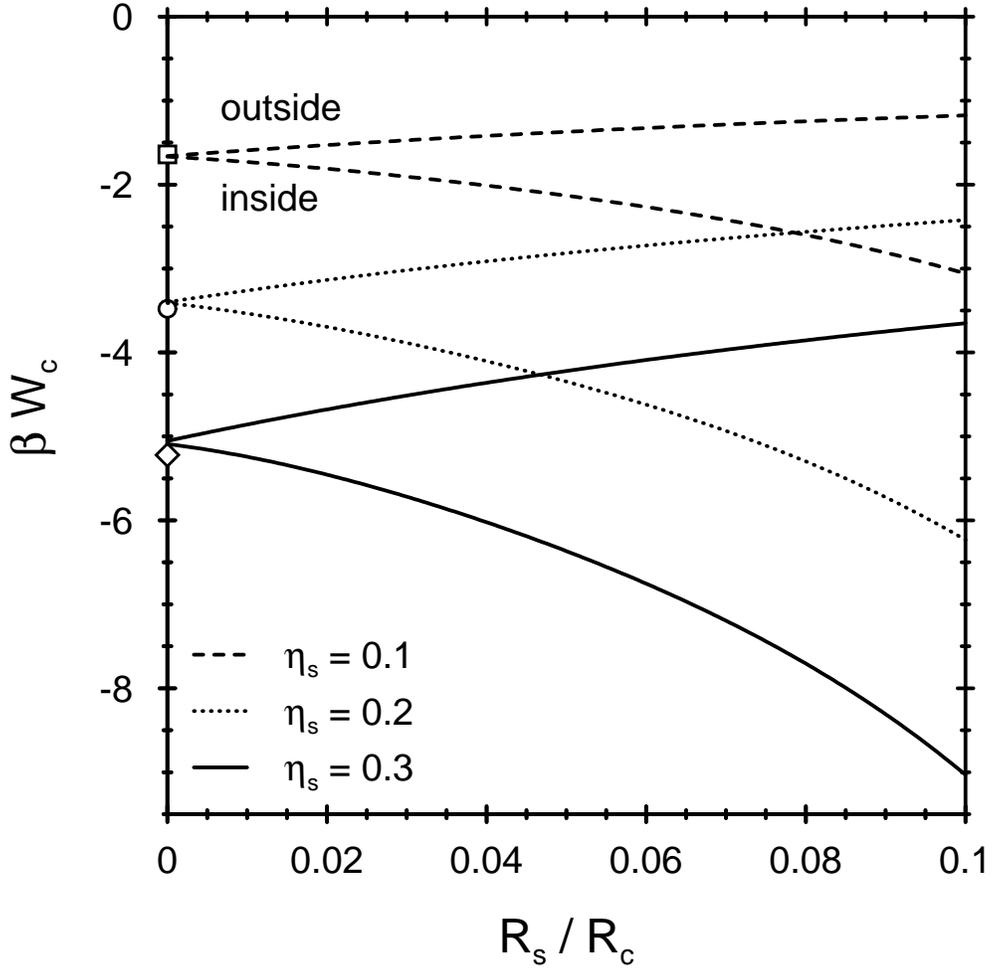,bbllx=10,bblly=60,bburx=550,bbury=580,width=13.5cm}

\vspace{0.5cm}

\caption{\label{fig:W0} Depletion Potential at contact $\beta W_{c}$ (see
Fig.~\ref{fig:pot}) as function of the cavity curvature $(R_{c}/R_{s})^{-1}$
for three packing fractions $\eta_{s}$ outside (upper curves) and inside
(lower curves) of the cavity with $R_{b}=5 R_{s}$. The symbols at $R_{c}^{-1}=0$ 
indicate simulation data for a flat wall \protect\cite{Dickman97}.}
\end{figure}

\begin{figure}
\centering\epsfig{file=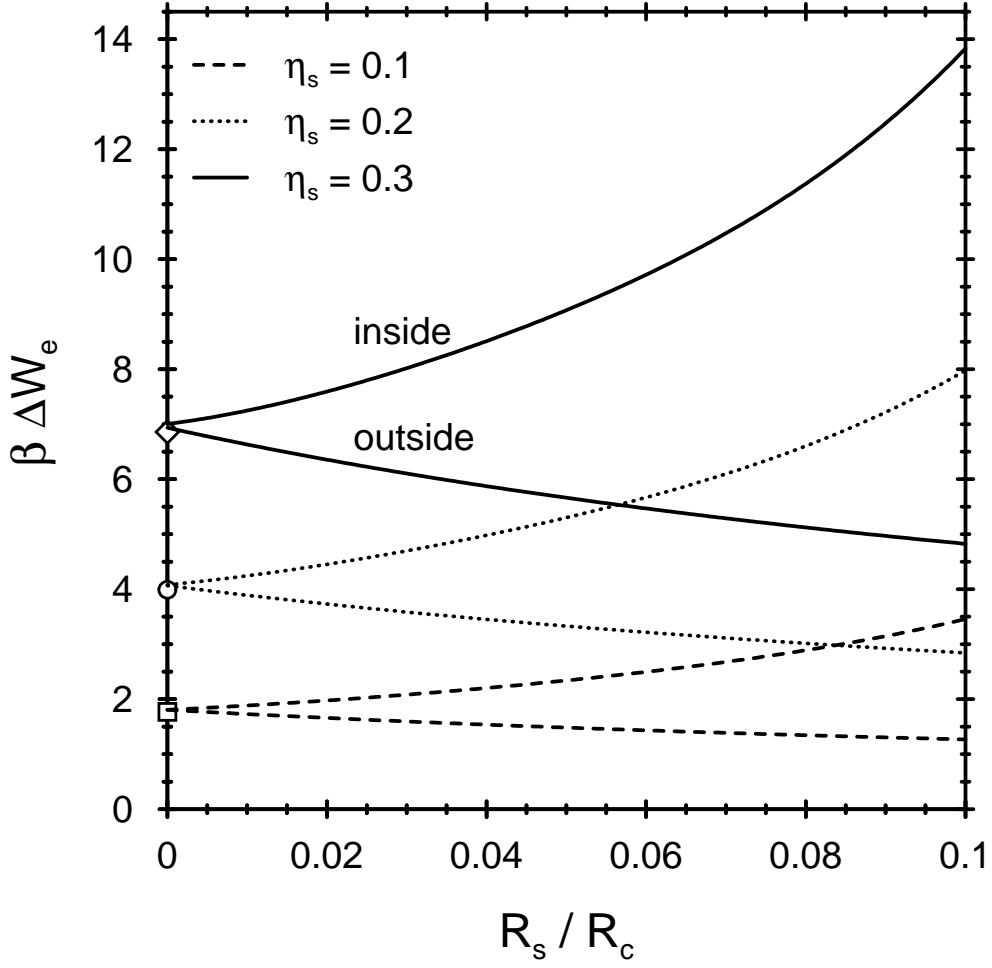,bbllx=10,bblly=60,bburx=550,bbury=580,width=13.5cm}

\vspace{0.5cm}

\caption{\label{fig:dW} Escape potential barrier $\beta \Delta W_{e}$ for a
big sphere with radius $R_{b}=5 R_{s}$ inside (upper curves) and outside
(lower curves) of a spherical cavity of radius $R_{c}$ for three values of
$\eta_{s}$. The symbols at $R_{c}^{-1}=0$ indicate simulation data for a flat
wall \protect\cite{Dickman97}.}
\end{figure}

In Figs.~\ref{fig:W0} and \ref{fig:dW} we discuss as function of the packing
fraction $\eta_{s}$ the relevance of the cavity curvature for the depletion
potential relative to the case of a planar wall, i.e., $R_{c}\to\infty$ by
focusing on the potential at contact $W_{c}$ and the escape potential
barrier $\Delta W_{e}$, respectively. As expected in the limit
$R_{c}\to\infty$ both quantities reach a common value for the outside and
inside potential. These common values are in very good agreement with
simulation data for the planar wall \cite{Dickman97}. For all values of $R_{c}$ the
absolute values of the potential parameters inside are lager than
outside. They increase stronger than linearly with increasing $\eta_{s}$. For
increasing curvature $1/R_{c}$ the outside potential becomes weaker whereas the
inside potential becomes much stronger; this holds for all values of
$\eta_{s}$. The gain of strength of the inside potential upon increasing the
cavity curvature is much more pronounced than the corresponding loss of
strength on the outside. This difference between the behavior inside and
outside widens strongly with increasing $\eta_{s}$. The dependence of the
depletion potential on the cavity curvature is surprisingly strong even for
large values of $R_{c}/R_{s}$. For $\eta_{s}=0.3$ and $R_{c}=50 R_{s}$ the
density profile $\rho_{s}(r)$ of the solvent without the big sphere near the 
curved wall differs only slightly from that near a flat wall; its contact value 
differs by
$1.5\%$ from the corresponding flat one. However, the potential at contact
still differs by almost $10\%$ from the corresponding flat value. This
amplification of the influence of the curvature can be understood in terms of
the geometric considerations of the overlap volumes leading to
\begin{equation}\label{Asakura:potential} 
\beta W_{c}^{AOa}(R_c)=-\eta_{s}\left(1+\frac{3 s R_{c}}{R_{c}+\gamma R_{b}}\right)
\end{equation}
with $s=R_{b}/R_{s}$ and $\gamma=+1$ and $-1$ for outside and inside,
respectively. Within AOa the absolute value of the amplitude of the first curvature
correction is the same inside and outside as can be seen from
\begin{equation}
\frac{\beta W_{c}^{AOa}(R_{c})}{\beta W_{c}^{AOa}(\infty)}=
1-\frac{3 \gamma s^{2}}{3 s+1}
\frac{1}{R_{c}/R_{s}}+{\cal O}\left( (R_{c}/R_{s})^{-2}\right).
\end{equation}
For $R_{b}\gg R_{s}$, i.e., $s\gg 1$ the curvature dependence of $W_{c}$ is
significantly enhanced.

\begin{figure}
\centering\epsfig{file=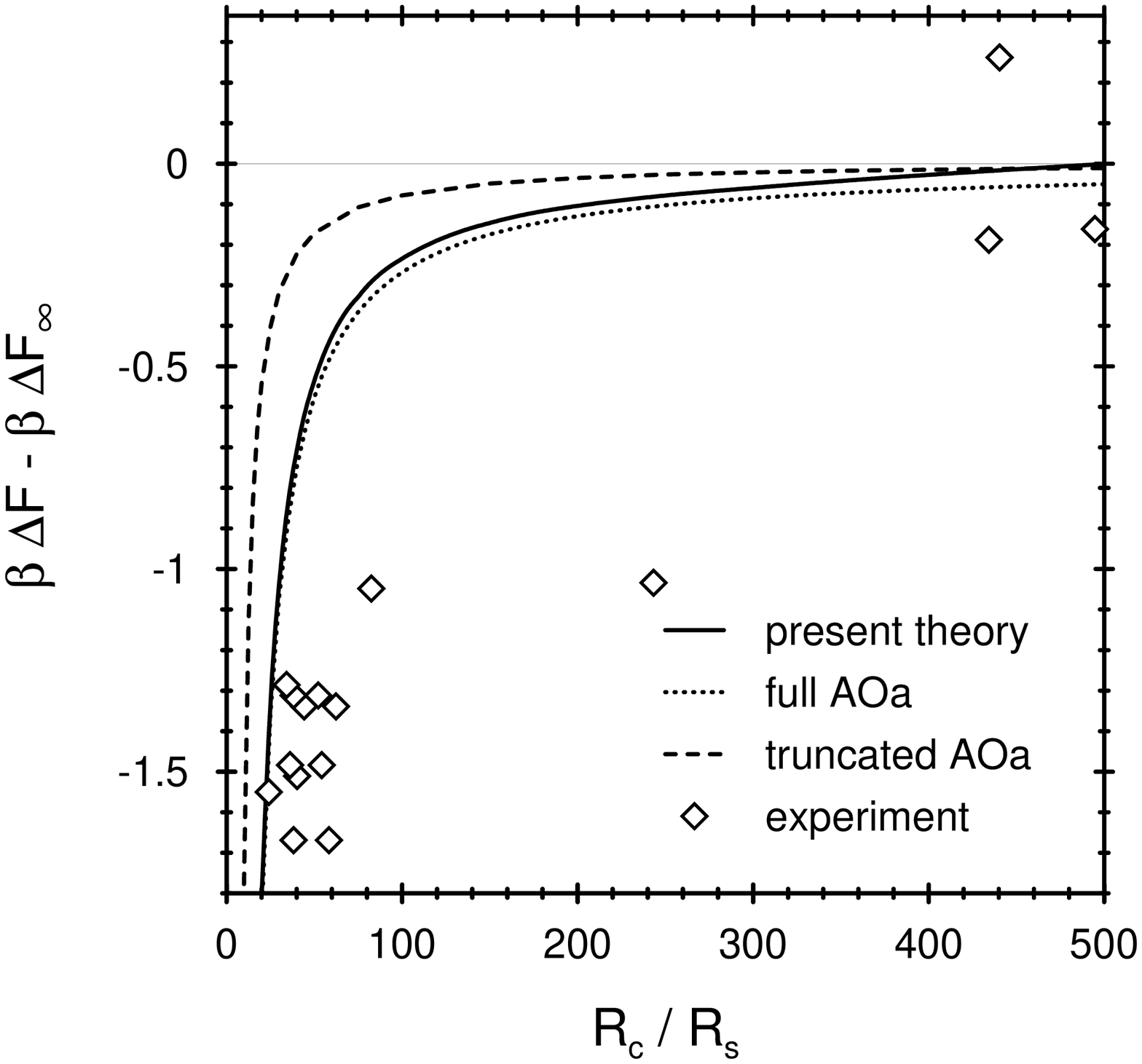,bbllx=10,bblly=60,bburx=550,bbury=580,width=13.5cm}

\vspace{0.5cm}

\caption{\label{fig:exp} Comparison of the experimental data 
\protect\cite{Dinsmore98} for $\beta \Delta F-\beta \Delta F_{\infty}$
(Eq.~(\protect\ref{prob})) as function of the local radius of curvature $R_c$
in units of $R_s$ with the prediction of the present theory (full line in
Fig.~\protect\ref{fig:pot}) for this quantity. The dotted line corresponds to the 
dotted line in Fig.~\protect\ref{fig:pot}. The dashed line denotes 
$\beta W_c^{AOa}-\beta W_{c,\infty}^{AOa}$ which are the AOa values for
$W(r;R_c)$ at wall contact which served in Ref.~\protect\cite{Dinsmore98} as
theoretical comparison.}
\end{figure}

In a recent experiment \cite{Dinsmore98} Dinsmore et al. used video microscopy
to monitor the position of a single big colloid particle with $R_b=5.7 R_s$ 
immersed in a solution of small colloids with $\eta_s=0.3$ inside a vesicle. 
These were charge stabilized colloids which are supposed \cite{Dinsmore98} to
resemble closely the
model system of hard spheres confined by a hard wall. From the probability
$p({\bf x})$ of finding the big colloid at the position ${\bf x}$ -- the quantity
actually measured in the experiment -- one can infer the depletion potential 
${\cal W}({\bf x})$ according 
to $p({\bf x}) = p_{bulk} e^{-\beta {\cal W}({\bf x})}$ with ${\cal W}({\bf x})=0$
in the bulk. Thus the spatial resolution of the experimentally determined
${\cal W}({\bf x})$ reflects that of $p({\bf x})$, which at best was approximately 
$2 R_s$ in Ref.~\cite{Dinsmore98}. 
In the experiment the big colloid particle was observed to be most of the time very
close to the vesicle wall inside a shell whose width was chosen to be $6.7 R_s$,
i.e., even larger than the optimal resolution $2 R_s$. Approximating
${\cal W}({\bf x})$ by $W(r;R_c)$, where $r$ is the minimal distance between
${\bf x}$ and the vesicle wall and $R_c$ the local radius of the wall at this
closest point (see below), the experimentally determined depletion potential
therefore corresponds to an average of $W(r;R_c)$ shown in Fig.~\ref{fig:pot}
over all visible oscillations.
Within this shell the big colloid was observed to be more often in regions with
a small local radius of curvature. This is in qualitative agreement with both
the AOa and our present approach.
Quantitatively, in Ref.~\cite{Dinsmore98} the ratio $p_{shell}/p_{bulk}$ of the
probabilities of finding the big particle within the shell and in the bulk,
respectively, was determined. The authors have documented the logarithm of this
ration, denoted as
\begin{equation}\label{prob}
\beta \Delta F :=- \ln \frac{p_{shell}}{p_{bulk}} \simeq
-\ln\left( \int_{V_{shell}} d^3 r e^{-\beta W(r;R_c)} / V_{shell} \right),
\end{equation}
relative to its value $\beta \Delta F_{\infty}$ near a flat wall, i.e., for 
$R_c\to \infty$;
$V_{shell}$ is the volume of the shell. In Fig.~\ref{fig:exp} we compare the
published data for $\beta \Delta F - \beta \Delta F_{\infty}$ with
our prediction for this quantity based on the actual
depletion potential as shown in Fig.~\ref{fig:pot}, with the corresponding
prediction if in Eq.~(\ref{prob}) and for $\beta \Delta F_{\infty}$ the full 
AOa is used, and with the so-called truncated AOa, 
$\beta W_{c}^{AOa}-\beta W_{c,\infty}^{AOa}$, i.e., the AOa values for
$W(r;R_c)$ at wall contact, which was used for comparison with theory
in Ref.~\cite{Dinsmore98}. We find that our theoretical prediction as well as
the full AOa are closer to the experimental data than the truncated AOa.
Given the high accuracy
of our calculation the remaining discrepancy cannot be due to insufficient
theoretical knowledge of $W(r;R_c)$. The fact that in Ref.~\cite{Dinsmore98}
only the radius of curvature in plane could be determined, polydispersity of the
solvent, and the possible presence of dispersion forces are among the candidates
for explaining this discrepancy. The small difference between the full AOa and
our theoretical results in Fig.~\ref{fig:exp} demonstrates that the present
spatial resolution cannot discriminate between the rich structure shown in
Fig.~\ref{fig:pot} and its AOa. This emphasizes the need for future experiments
with significantly increased spatial resolution.

For a cavity of general shape \mbox{$z=f({\bf R}=(R_x,R_y))$} of its surface relative to
a suitable chosen $(x,y)$ reference plane our results allow us to determine
approximately {\em lateral} depletion forces. To this end we introduce normal
coordinates \cite{Zia85} $(R_x,R_y,r)$ such that 
${\bf x}=(x,y,z)=(R_x,R_y,f({\bf R}))+r {\bf n}({\bf R})$ where $r$ is the minimal
distance of the point ${\bf x}$ from the surface $f({\bf R})$, $(R_x,R_y)$
($\neq (x,y)$) are the lateral coordinates of that point on the surface closest
to ${\bf x}$, and ${\bf n}=(-\nabla f({\bf R}),1)/\sqrt{1+(\nabla f)^2}$ is 
the local
surface normal pointing towards ${\bf x}$. The actual depletion potential
${\cal W}({\bf x};[f])$, which depends functionally on $f({\bf R})$, can be
approximated by ${\cal W}({\bf x};[f])\simeq W(r({\bf x});R_c({\bf R}({\bf x})))$.
Here we have assumed that the cavity surface varies sufficiently smoothly so that
its local two principal radii of curvature can be described by the single radius
$R_c({\bf R})$. Within this approximation the components $F_{lat}^{(i)}({\bf x})$ 
of the lateral depletion force in the direction of the tangential vector
${\bf t}_i$, $i=1,2$, is given by
\begin{equation} \label{lateral}
F_{lat}^{(i)}({\bf x})=\frac{\partial W(r;R_c({\bf R}))}{\partial R_c} 
\sum_{j=x,y} \frac{\partial R_c({\bf R})}{\partial R_j} {\bf t}_i \cdot 
\nabla R_j~,
\end{equation}
with $r=r({\bf x};[f]), {\bf R}={\bf R}({\bf x};[f]),\nabla=\partial/\partial {\bf x}$,
and ${\bf n} \cdot {\bf t}_i=0={\bf t}_1 \cdot {\bf t}_2$. 
The factor $\partial W/\partial R_c$ can be determined from our above results.
Its absolute value increases with decreasing radius of curvature. Its sign 
differs inside and outside of the cavity and changes as function of the
normal distance $r$. A big sphere approaching the convex nonspherical wall from
inside (outside) is exposed to an {\em oscillating} lateral force, which pulls
the sphere towards regions of the surface with a small local radius of
curvature close to a minimum (maximum) of the depletion potential and
pushes it away from these regions close to a maximum (minimum). These 
oscillations of the lateral force originate from the packing effects of the 
small spheres.  When the big sphere has reached
the wall the lateral force pulls it along the surface to that point
with the smallest local radius of curvature inside the cavity and pushes 
it away from this point towards the point with the largest local radius of 
curvature outside
of the cavity, respectively. This can already be understood within the AOa 
which predicts 
$\partial (\beta W_c^{AOa}(R_c))/\partial R_c = -3 \eta_s \gamma s R_b
(R_c+\gamma R_b)^{-2}$ (see Eq.~(\ref{Asakura:potential})).
Since Eq.~(\ref{Asakura:potential}) provides a very good description of the 
contact value of the depletion potential, even for high packing fractions, the
above expression for its derivative is also quantitatively reliable.

Applying Eq.~(\ref{lateral}) for an ellipsoidal cavity with semi-axes 
$a=b=20 R_s$
and $c=30 R_s$ and for a big sphere with $R_b=5 R_s$ immersed in a fluid of small
spheres with $\eta_s=0.3$ yields at contact a maximal ratio of the lateral 
force to the normal
force of $|F_{lat}/F_{norm}|=0.07$ inside and outside of the cavity.
Such quantitative predictions for the size of entropic lateral forces along
curved confining walls still await tests by simulations and experiments.

\onecolumn

\begin{references}
\bibitem{Zimmerman97} S.B. Zimmerman and A.P. Minton, Ann. Rev. Biophys. Biomol. 
Struct. {\bf 22}, 27 (1993); A.P. Minton, Curr. Opin. Biotechnol. {\bf 8}, 65 (1997).
\bibitem{Crocker94} See, e.g., J.C. Crocker and D.G. Grier,
  Phys. Rev. Lett. {\bf 73}, 352 (1994).
\bibitem{Dinsmore96} A.D. Dinsmore, A.G. Yodh, and D.J. Pine, Nature {\bf383}, 
239 (1996).
\bibitem{Kaplan94a} P.D. Kaplan, J.L. Rouke, A.G. Yodh, and D.J. Pine, Phys. Rev.
Lett. {\bf 72}, 582 (1994).
\bibitem{Kaplan94b} P.D. Kaplan, L.P. Faucheux, and A.J. Libchaber,
Phys. Rev. Lett. {\bf 73}, 2793 (1994).
\bibitem{Sober95} D.L. Sober and J.Y. Walz, Langmuir {\bf 11}, 2352 (1995).
\bibitem{Milling95} A. Milling and S. Biggs, J. Colloid Sci. {\bf 170}, 604 (1995).
\bibitem{Ohshima97} Y.N. Ohshima, H. Sakagami, K. Okumoto, A. Tokoyoda, T. Igarashi,
K.B. Shintaku, S. Toride, H. Sekino, K. Kabuto, and I. Nishio,
Phys. Rev. Lett. {\bf 78}, 3963 (1997).
\bibitem{Rudhardt98} D. Rudhardt, C. Bechinger, and P. Leiderer, Phys. Rev. Lett.
{\bf 81}, 1330 (1998).
\bibitem{Mao95} Y. Mao, M.E. Cates, and H.N.W. Lekkerkerker, Physica A {\bf222},
10 (1995).
\bibitem{Biben96} T. Biben, P. Bladon, and D. Frenkel, J. Phys.: Condens. Matter
{\bf 8}, 10799 (1996).
\bibitem{Goetz98a} B.G\"otzelmann, R. Evans, and S. Dietrich, Phys. Rev. E {\bf57},
6785 (1998).
\bibitem{Goetz98b} B. G\"otzelmann, Ph.D. thesis, Bergische Universit\"at Wuppertal
(1998).
\bibitem{Dickman97} R. Dickman, P. Attard, and V. Simonian, J. Chem. Phys.
{\bf 107}, 205 (1997).
\bibitem{RF89} Y. Rosenfeld, Phys. Rev. Lett. {\bf 69}, 980 (1989).
\bibitem{comment} Inside the cavity the chemical potential is adjusted such
that its value corresponds to the value for $\eta_s$ in unbounded space. For
small cavities one has to be aware of the difference between the depletion
potentials calculated by density functional theory corresponding to a grand
canonical ensemble and experimentally determined depletion potentials inside of
cavities corresponding to a canonical ensemble.
The results in A. Gonz\'alez, J.A. White, F.L. Rom\'an, 
S. Velasco, and R. Evans, Phys. Rev. Lett. {\bf 79}, 2466 (1997) indicate that
the difference between the ensembles is only relevant for cavity sizes in the order
of a few small sphere radii.
\bibitem{Dinsmore98} A.D. Dinsmore, D.T. Wong, P. Nelson, and A.G. Yodh,
Phys. Rev. Lett. {\bf 80}, 409 (1998).
\bibitem{Asakura54} S. Asakura and F. Oosawa, J. Chem. Phys. {\bf 22}, 1255 (1954).
\bibitem{Zia85} R.K.P. Zia, Nucl. Phys. B {\bf 251} [FS 13], 676 (1985). 
\end{references}
\end{document}